# Tunable phase diagram and vortex pinning in a superconductor-ferromagnet bilayer

L. Y. Zhu[1], Marta Z. Cieplak[1,2], and C. L. Chien[1]

[1]*Department of Physics and Astronomy, The Johns Hopkins University, Baltimore, MD 21218, USA*

[2]*Institute of Physics, Polish Academy of Sciences, 02 668 Warsaw, Poland*

We have observed the evolution of phase diagram and vortex pinning using a single ferromagnet/superconductor bilayer of $[Co/Pt]_8$/Nb through a special demagnetization procedure. It induces a continuous and reversible change of the domain width with equal positive/negative domains enabling the observation of the predicted tunable phase diagram. The tunable domain pattern also systematically affects vortex pinning. We have determined the dependence of the activation energy of vortex pinning on domain width, temperature, and magnetic field.

Superconducting(S)/ferromagnetic(F) bilayers (SFBs) have been important model systems to study the strong and intricate interplay between superconductivity and magnetism, two long-range orders that are mutually exclusive in the bulk [1-3]. In an SFB, the stray fields from the F-layer due to magnetic domains interact with the Cooper pairs in the S layer via the long-range (orbital) interaction. In SFBs, the F layers with perpendicular magnetic anisotropy (PMA) have unique attributes compared to those with in-plane anisotropy. The stray magnetic field that emanates from a magnetic domain penetrating into the S layer has a well-defined direction and can be systematically altered by an external magnetic field.

Both the nucleation of the superconducting order parameter (OP) and the pinning of vortices [1-3] are strongly affected by stray fields. Since superconductivity nucleates in regions where the local fields are minimal [2], stray field has a profound effect as manifested in the phase diagram consisting of $T_c(H)$-lines, which are the field dependence of the superconducting critical temperature $T_c$. Some aspects of the theoretical predictions [4] of the reentrant phenomena have been observed experimentally [5, 6]. The strong dependence due to magnetic stripe domain width has also been predicted [7], but not yet verified experimentally. Such studies would require a series of SFB's each with an identical S layer and a specific domain pattern in the F layer with PMA with *equal* amount of positive/negative domains and that domain width can be continuously tuned. This has not been previously accomplished.  For example, a magnetic field was previously [6] used to partially switch the multilayer [Co/Pt]$_n$ to create domain patterns with *unequal* amount of positive/negative domains with *unequal* domain width. In contrast to theoretically predicted *symmetric* phase diagram, the observed phase diagram is distinctively *asymmetrical* with respect to $H = 0$. Such partially switched system is also susceptible to magnetic field during the determination of the phase diagram.

Below $T_c$, the magnetic textures also create pinning potentials for vortices [1-3]. Experiments, realized mostly in ordered arrays of magnetic nanodots, have uncovered many novel effects, including the commensurability between the vortex lattice and the underlying dot array, vortex channeling, and other dynamic effects [3]. In comparison, fewer studies employ planar SFB's, which offer the prospects of reversible manipulation of the domain pattern. One notable example is SFB's with regular stripe domains, which may be aligned to produce anisotropic vortex pinning [8, 9]. Theories predict that pinning should be enhanced in SFB's [10], a prospect claimed by some experiments [2, 8, 11] but contradicted by others [12]. A comprehensive picture remains lacking.

In this work we study these phenomena in a single planar SFB consisting of Nb as the S layer and Co/Pt multilayer with PMA as the F layer. It has not been feasible to fabricate a series of SFBs with varying domains in the F layer with equal positive/negative domains and identical superconducting properties in the S layer. Instead, we use a single sample and tune the domain width $L$ by a novel demagnetization procedure, with which the specific stripe domain pattern with zero magnetization can be reversibly defined and erased, resulting in tuning the phase diagram as predicted theoretically [7]. Furthermore, we have also studied the influence of $L$ on the vortex pinning behavior. We have determined the dependence of vortex activation energy on $L$, $T$, and $H$, and provided a comprehensive picture of pinning in the SFB with randomly oriented stripe domain pattern.

The multilayer Pt(10)/[Co(0.6)/Pt(0.3)]$_8$/Si(10)/Nb(20)/Si(10), with thickness denoted in nanometers, was grown by sputtering at room temperature on a Si substrate. The central Si(10) layer has been inserted to ensure the absence of proximity effects between [Co/Pt]$_8$ and Nb. In this work, the magnetic field, denoted as $H$, has been applied perpendicular to the film plane. The [Co/Pt]$_8$ multilayer shows a square hysteresis loop with a coercive field of $H_C \approx 750$ Oe at 10 K as shown in Fig. 1a. The remnant state (denoted as sF) has a single domain with a saturated magnetization $M_S$. Using ac demagnetization, the F layer can be made unmagnetized with $M = 0$ at $H = 0$, where [Co/Pt]$_8$ breaks into equal amount of up and down stripe domains with equal domain width. Upon increasing $H$ along the initial (red) curve, $M$ remains very small until $H$ approaches that of $H_C \approx 750$ Oe. In the field range of $|H| \leq 400$ Oe, the magnetic domains remain essentially unchanged, and the external field simply alters the stray field entering the S layer above a magnetic domain.

Previously demagnetization method with either perpendicular or in-plane field has been used to create different domain patterns [13, 14]. Unique to this work is the use of a novel demagnetizing method, in which an ac magnetic field of decreasing magnitude is applied at an angle $\theta$ with respect to the film plane as schematically shown in the inset of Fig. 1i. Magnetic force microscopy (MFM) images (Fig. 1c-h) reveal meandering stripe domains with equal amounts of positive and negative domains, but the average domain width increases monotonically with increasing $\theta$. The meandering stripes in Co/Pt multilayer minimize magnetostatic energy. A two-dimensional Fourier analysis of the MFM images gives a Gaussian distribution with a mean domain period $2L$ and a standard deviation $\Delta L$. The domain size $L(\theta)$ increases from 328 nm at $\theta = 0°$ to 520 nm at $\theta = 90°$ with a standard deviation $\Delta L$ of less than 8% (Fig. 1i). The $L(\theta)$ dependence for ferromagnets with PMA can be qualitatively understood as follows. While the magnetic moment is out-of-plane in the uniform domain area, there is an in-plane component for the magnetic moments within the domain walls. Demagnetizing at a smaller $\theta$ tends to align the spins in plane, create more domain walls, hence reducing the domain width. This novel demagnetizing technique allows the realization of a wide range of domain width $L(\theta)$ using a *single* sample.

The electrical resistance $R$ of the SFB, has been measured on sample 2 mm x 5 mm with four contact pads along the long direction and 2 mm between the voltage leads. After each field processing at different $\theta$ as well as that in the sF state, $R$ has been measured in a perpendicular magnetic field $H$ at temperature just below $T_c$ in increments of 2-3 mK. As shown in Fig. 2a-d, while $R$ (normalized to $R_N$, the normal-state value at 10 K) increases with increasing $H$ and $T$, the field dependence for different $\theta$ is markedly different. The value of $R$ shows one sharp minimum at $H = 0$ for the sF state. The minimum broadens for $\theta = 0°$, evolves into two shallow minima at about ± 150 Oe for $\theta \approx 15°$, and finally becomes two deep minima for $\theta \approx 90°$. Defining $R/R_N = 0.5$ as the representative temperature for superconductivity, the values of $T_c(H)$ can be extracted from Fig. 2a-d. As shown in Fig. 2e, $T_c(H)$ is linear in $H$ for the sF state, which is single-domain and without stray field. Using this result and the coherence length in the dirty limit [15], $\xi(T) = \xi(0)/(1-T/T_{c0})^{1/2}$, and the upper critical field $H_{c2}(T) = \Phi_0/2\pi\xi^2(T)$, where $\Phi_0 = 20.68$ G·$\mu m^2$ is the flux quantum, we obtained $T_{c0} = 7.95$ K and $\xi(0) \approx 12.0$ nm.

As shown in Fig. 2e, $T_c(H)$ is non-linear for the states processed at $\theta$. Although $T_c(H)$ has a single maximum at $H = 0$ for small $\theta$, it changes to bimodal at large $\theta$ with two maxima $T_c^{max}$ at $H \approx \pm 150$ Oe, revealing the reentrant phenomenon [6]. The physical origin of the bimodal $T_c(H)$ with $T_c^{max}$ is schematically depicted in Fig. 1b. The local field is $H_{loc} = H + b_z$, a sum of external field $H$ and stray field $b_z$ with the same direction as that of the domain. At $H = 0$, $H_{loc} = b_z$ with the same magnitude but opposite direction for the up and the down domains. Superconductivity nucleates near the domain wall regions (blue areas in Fig. 1b). The external field $+H$ applied perpendicular to the plane increases (decreases) the stray field in the up (down) domain. At $T_c^{max}$, $H_{loc} = 0$ when $H$ compensates the stray field and superconductivity is induced above the oppositely oriented magnetic domains.

The highest $T_{c0} = 7.95$ K occurs at $H = 0$ in the sF state which has effectively infinite $L$, whereas $T_c$ is suppressed in all other cases (Fig. 2e). As shown in Fig. 2f, there are two distinct behaviors for $T_c(H)$. For small $\theta$ ($L \leq 337$ nm), $T_c(H)$ shows one maximum $T_c(0)$ at $H = 0$, whose values decreases with increasing $L$. For larger $\theta$ ($L > 337$ nm), $T_c(H)$ is bimodal with two maxima at $T_c^{max}$, whose values increase with $L$ and appear to saturate at large $L$. The demarcation point in Fig. 2f is at $T_c = 7.905$ K with $L = 337$ nm, corresponding to $2.12\xi$, which is close to $2\xi$. This key result affirms that $L \approx 2\xi$, i.e., the half-period $L$ is close to that of the OP.

The OP nucleation in the SFB with periodic stripe domains has been considered theoretically [7] in the context of local field $H_{loc} = H + b_z$. The phase boundary $T_c(H)$ has been found to depend on the relative size of $L$ and $\xi$. At $L \ll \xi$, the OP extends over several domains so that the spatial variation of the stray field is averaged out resulting in a single $T_c$ maximum at $H = 0$ as we have experimentally realized in states with $L \leq 337$ nm. We note $L \approx 2\xi$ in our case and not $L \ll \xi$. At intermediate $L$, reentrant $T_c(H)$-line appears with two maxima at $H \approx b_z$, as we have observed in states with $L > 337$ nm.

Theory also considers the case for very large $L \gg \xi$, where $T_c(H)$ remains bimodal but with a cusp at $H = 0$ in the $T_c(H)$-line. The largest $L$ of about 520 nm may not be sufficiently large to be in this limit. Previously features of reentrant phase diagram have been observed in isolation. We have been able to observe its *continuous* evolution because of our unique sample.

We next turn to the vortex pinning in the tunable SFB. In Fig. 3a, we show the $H$-dependence of low $R$ at $R/R_N \sim 10^{-2}$-$10^{-3}$ measured at a lower temperature of 7.882 K ($T/T_{c0} = 0.99$). This variation of resistance is the result of the activation of pinned vortices [16-18]. For all angles $\theta > 13°$ we see two minima in $R/R_N$ located symmetrically at about ±100 Oe with a maximum at $H = 0$, which develops into a plateau for $\theta = 45°$ and 90°. In Fig. 3b we show the critical current $I_c(H)$, defined by the criterion of voltage exceeding $10^{-6}$ V over 2 mm between the voltage leads, extracted from the *I-V* characteristics measured at 7.882 K. For the sF state (open squares), $I_c(H)$ decreases sharply and monotonically with $H$, whereas that of the state with $\theta = 90°$ shows a peak at $H \approx 100$ Oe. The presence of domains enhances $I_c$ at 100 Oe, but strongly suppresses $I_c$ for $H \leq 50$ Oe, each by about an order of magnitude. At lower $T$ (at $T/T_{c0} = 0.98$ and 0.89), the magnitude of $I_c$ is much larger. But as also shown in Fig. 3b, $I_c$ reaches saturation as $H$ approaches zero.

To understand this pinning behavior we examine $R$ as a function of $1/T$ as shown in Fig. 3c and 3d. At sufficiently low $T$ all curves become linear, exhibiting a clear Arrhenius behavior. In the case of the sF state (Fig. 3c) the |slope| of the linear dependence decreases monotonically with increasing $H$, indicating a continuous decrease

of the flux activation energy $U$. However, the results for the state with $\theta = 90°$ are very different (Fig. 3d). The data for $0 \leq H \leq 40$ Oe are nearly indistinguishable from each other. The |slope| increases with increasing $H$ for 50 Oe $\leq H \leq$ 100 Oe, above which the |slope| decreases. To extract the activation energy $U$ we follow the previous analysis of the flux activation regime [18] and fit the linear portions on the Arrhenius plots by the dependence of $\ln(R/R_N) = -U_0(H)/(k_B T) + K(H)$. The activation energy in general is $U(T, H) = U_0(H) - K(H)T$, where, $U_0(H)$ is the zero-temperature activation energy, and $K(H)$ is the coefficient in the linear $T$ correction. Since $U$ vanishes at $T_c$, it follows that $K(H) = U_0(H)/T_c(H)$. This is indeed the case as all the data falls on a straight line of $U_0(H)/T_c(H)$ vs. $K(H)$. Thus, $U$ has the form of $U(T,H) = U_0(H)[1 - T/T_c(H)]$.

In Fig. 4a, $U_0(H)$, the slope of $\ln(R/R_N)$ vs. $1/T$, is shown for various states of the F-layer. In the sF state, $U_0(H)$ varies as $\ln H$, which has been reported in epitaxial 2-nm Nb film [18] and attributed to a distribution of pinning strengths, and the preferential "filling" of the strongest pinning sites [19, 20]. The $U_0$ in our 20 nm thick Nb film is about an order of magnitude larger than that for 2 nm thick Nb [18], consistent with pinning in the 2D limit, where $U_0$ is proportional to the film thickness.

The zero-temperature activation energy $U_0$ is enhanced by the presence of domains, most prominent at $H \leq 40$ Oe. As shown in Fig. 4a, the enhancement over that of sF is the largest for $\theta = 0°$, which also retains partly the $\ln H$ dependence. With increasing $\theta$, $U_0$ gradually reduces and becomes $H$ independent, which we will comment later. At intermediate $H$ ($40 \leq H \leq 150$ Oe), $U_0$ takes the form of peaks. At $H \geq 150$ Oe, $U_0$ rapidly reduces for all $\theta$, and approaches that of the sF state, when the high vortex density exceeds that of domain-induced pinning centers. In Fig. 4b, we show $U_0$ in the intermediate $H$-region on a linear scale and vertically shifted for clarity. One observes peaks at $H_1 \approx 40$ Oe for $\theta = 0°$, at $H_2 \approx 70$ Oe for $\theta = 10°$, 13°, 15°, and at $H_3 \approx 90$ Oe for $\theta \geq 15°$. Both $H_2$ and $H_3$ are present for $\theta = 15°$. For $\theta \geq 20°$ only $H_3$ peak remains.

The peaks in $U_0$ indicate enhanced pinning when the vortex lattice becomes commensurate with the pinning potential created by the domain pattern. The first matching field $H$, a condition when a single flux quantum is pinned by each pinning center, corresponds to the average area per vortex $A = \Phi_0/H$ [3]. In SFBs with stripe domains, since negative domains are anti-pinning, vortices exist only in half of the area, $A_n/2 = \Phi_0/2H_n$, equal to $2.6 \times 10^{-9}$ cm$^2$, $1.5 \times 10^{-9}$ cm$^2$, and $1.1 \times 10^{-9}$ cm$^2$, for $H_1$, $H_2$, and $H_3$, respectively. Although the disorder in SFB would limit the matching areas, leading to broadening and/or shifting of the matching fields, we nevertheless gain some insight by considering the pinning for the regular domain pattern with a period $2L$ as $L$ increases.

When $L$ is small, a single 1D vortex chain with the vortex-vortex separation $a$ is pinned by each domain. The average area per vortex is $aL = \Phi_0/2H$. Shown in Fig.4c is $a$ calculated using the observed values of $H_1$, $H_2$ and $L$. The inset shows vortex-domain configurations at $H_1$ (red) and $H_2$ (blue). Interestingly, $a=773$ nm at $H_1$ is identical to that of the Abrikosov lattice constant in continuous film, $a^2 = (4/3)^{1/2} \Phi_0/H_1$, suggesting the vortex chains are correlated across neighboring domains to form Abrikosov lattice (red inset). In the large $L$ limit, the vortex chains form Abrikosov lattice *within* each domain with $a^2 = (4/3)^{1/2} \Phi_0/2H_3$, from which we obtain $a = 364$ nm using $H_3 = 90$ Oe as shown in Fig.4c by the horizontal line. In between is the transition region exemplified by $\theta = 15°$, where $H_3$ peak appears along with $H_2$, sharing the same $H_3$ for $\theta > 15°$. This signals the formation of double vortex chains inside domains, separated by a distance $h < L$ (green configuration in the inset). The average area per vortex is $bL/2 = \Phi_0/2H_3$, where $b$ is the separation along each chain, and $a$ is defined by $a^2 = (b/2)^2 + h^2$. We estimate $h$ for $\theta = 15°$ by noticing that $a$ in the 1D chain at $H_2$ should be comparable to $a$ in double chain at $H_3$, which gives $h = 0.6L$. Using this value we calculate $a$ for intermediate $L$. The points calculated with $h = 0.6L$ run through the horizontal Abrikosov line.

The distribution of $L$ in the SFB results in variation of the local vortex patterns. However, with the exception for $\theta = 15°$, for every other $\theta$ there is only one matching field, suggesting that the configurations discussed above are quite robust. The matching conditions consistent with the vortex chains have been observed in the case of Ni wires on Nb films [21]. Recently, a direct imaging by the scanning tunneling microscopy has revealed the formation of multiple vortex chains in $NbSe_2$ with Py on the top [9].

As shown in Fig. 4a, with increasing $\theta$, the suppression of $U_0$ at $H \leq 40$ Oe increases. This is due to the local stray field as shown in Fig. 1b. At the center of the domain, the stray field scales as $1/L$, hence $U_0$ at $H \approx 1$ Oe follows a similar dependence. In addition, the confinement of vortices by wider domains is less effective; single vortices are freer to flow until the vortex density reaches the matching value. This results in the $H$-independent $U_0$ at $H \leq 40$ Oe for large $L$.

Finally we mention the finite $T$ effects. The activation energy $U_0$ at $T = 0$ for the sF state is the lowest as shown in Fig. 4a. At finite $T$, $U(T,H) = U_0(H)[1- T/T_c(H)]$ such that at $T = 7.882$ K and for $H \leq 50$ Oe, $U(T,H)$ for sF is larger than that for $\theta = 90°$ as shown in Fig. 4d. This accounts for both the suppression of the critical current and the plateau in the resistance observed for $H \leq 50$ Oe.

In conclusion, we have used a novel demagnetization procedure to tune the width of domains in a single planar superconducting/ferromagnetic bilayer (SFB). This allows us to continually tune the phase boundary from that of a single maximum to the reentrant one. By extracting the activation energy for vortex pinning in the tunable SFB, we have made a comprehensive study of vortex pinning on domain width, temperature and magnetic field in the structure with randomly oriented stripe domains.

This work was supported by NSF Grant DMR05-20491, and by Polish MNiSW grant N202 058 32/1202.

**References:**

[1] M. V. I. F. Lyuksyutov and V. L. Pokrovsky, Adv. Phys. **54**, 67 (2004).

[2] A. Yu. Aladyshkin *et al.*, Topical review, Supercond.Science and Technol. **22**, 053001 (2009), and references within.

[3] M. Velez *et al.*, Topical Review, J. Mag. Magn. Mat. **320**, 2547 (2008), and references within.

[4] A. I. Buzdin and A. S. Mel'nikov, Phys. Rev. B **67**, 020503(R) (2003); A. Yu. Aladyshkin, *et al.*, Phys. Rev. B **68**, 184508 (2003).

[5] Z. Yang *et al.*, Nature Mat. **3**, 793 (2004).

[6] W. Gillijns *et al.*, Phys. Rev. Lett. **95**, 227003 (2005), and W. Gillijns *et al.*, Phys. Rev. B. **76**, 060503(R) (2007).

[7] A. Yu. Aladyshkin and V. V. Moshchalkov, Phys. Rev. B **74**, 064503 (2006).

[8] V. Vlasko-Vlasov *et al.*, Phys. Rev. B **77**, 134518 (2008); *ibid.*, **78**, 214511 (2008); A. Belkin, *et al.*, Phys. Rev. B **77**, 180506(R) (2008).

[9] G. Karapetrov *et al.*, Phys. Rev. B **80**, 180506(R) (2009).

[10] L. N. Bulaevskii, *et al.*, Appl. Phys. Lett. **76**, 2594 (2000); Yu. I. Bezpyatykh *et al.*, Sov. Phys. -Solid State **43**, 1827 (2001).

[11] M. Z. Cieplak *et al.*, J. Appl. Phys. **97**, 026105 (2005); M.Z. Cieplak *et al.*, Phys. Stat. Sol. (c) **2**, 1650 (2005).

[12] M. Feigenson, *et al.*, J. Appl. Phys. **97**, 10J120 (2005).

[13] O. Hellwig, *et al.*, J. Mag. Magn. Mat. **319**, 13 (2007).

[14] L. Y. Zhu, T. Y. Chen and C. L. Chien, Phys. Rev. Lett. **101**, 017004 (2008).

[15] M. Tinkham, Introduction to superconductivity, Dover Publications 2004.

[16] T.T.M. Palstra *et al.*, Phys. Rev. B **41**, 6621 (1990).

[17] O. Brunner *et al.*, Phys. Rev. Lett. **67**, 1354 (1991).

[18] J. W. P. Hsu and A. Kapitulnik, Phys. Rev. B **45**, 4819 (1992).

[19] M. Inui *et al.* Phys. Rev. Lett. **63**, 2421 (1989).

[20] S. Martin and A. F. Hebard, Phys. Rev. B **43**, 6253 (1991).

[21] D. Jaque *et al.* Appl. Phys. Lett. **81**, 2851 (2002).

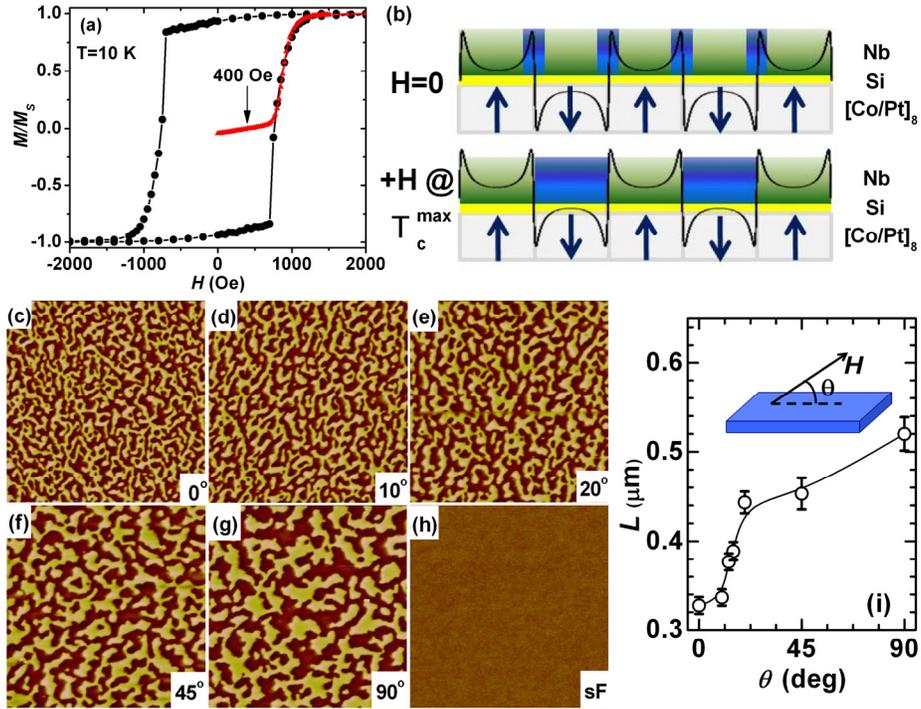

FIG. 1: (a) Hysteresis loop at 10 K of [Co(0.6)/Pt(0.3)]$_8$ in a perpendicular field $H$ with the initial curve shown in red, (b) Schematics of stray field in Nb due to stripe domains in [Co(0.6)/Pt(0.3)]$_8$ (c)-(h) MFM images (10 x 10 $\mu m$) at 300 K after demagnetization at angle $\theta$ = 0°, 10°, 20°, 45°, 90°, and sF respectively with the resultant average domain half period $L$ shown in (i).

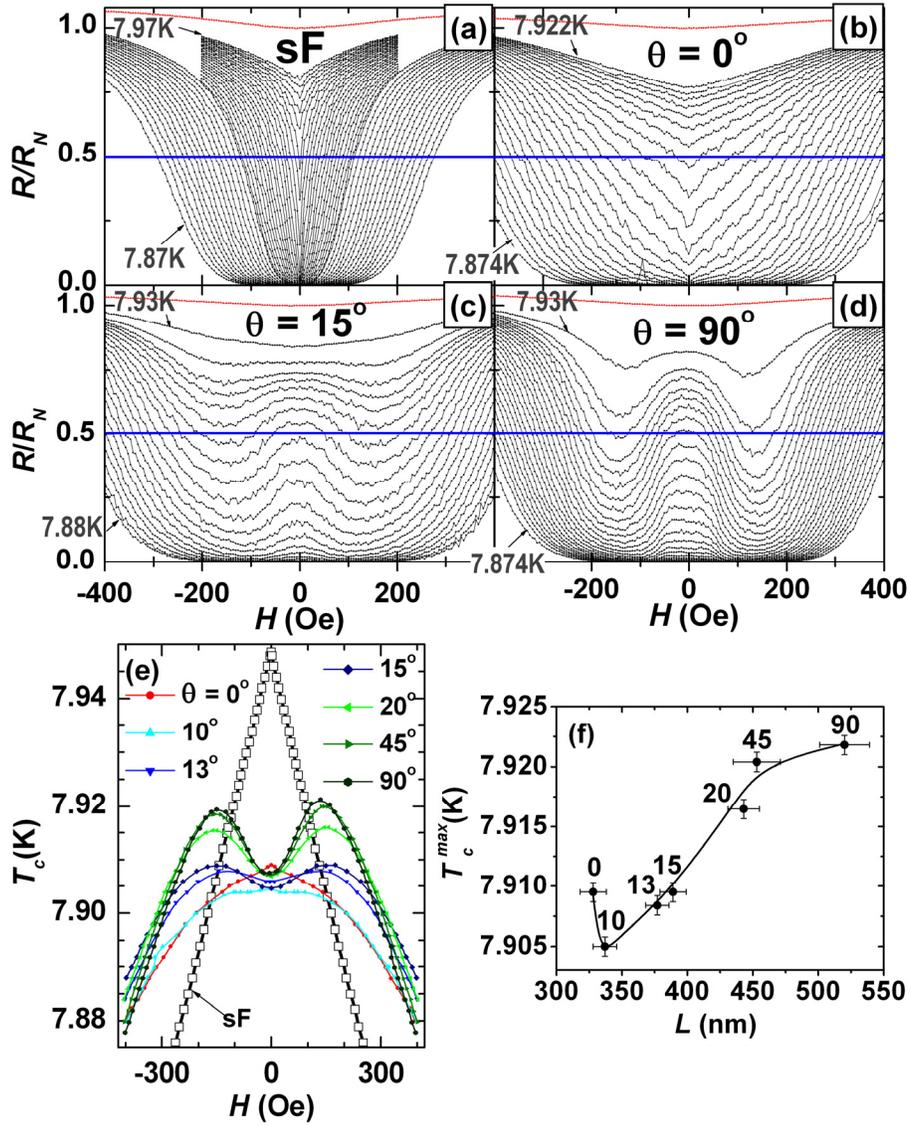

FIG. 2: Resistance $R$ (normalized to $R_N$, the value at 10 K shown in red) versus $H$ in increments of 2-3 mK for states of (a) saturated sF, and demagnetized at (b) $\theta = 0°$ (c) 15° and (d) 90°. (e) Values of $T_c$ at $R/R_N = 0.5$ for different $\theta$, and (f) the maximal $T_c$ as a function of domain half period $L$ in (e).

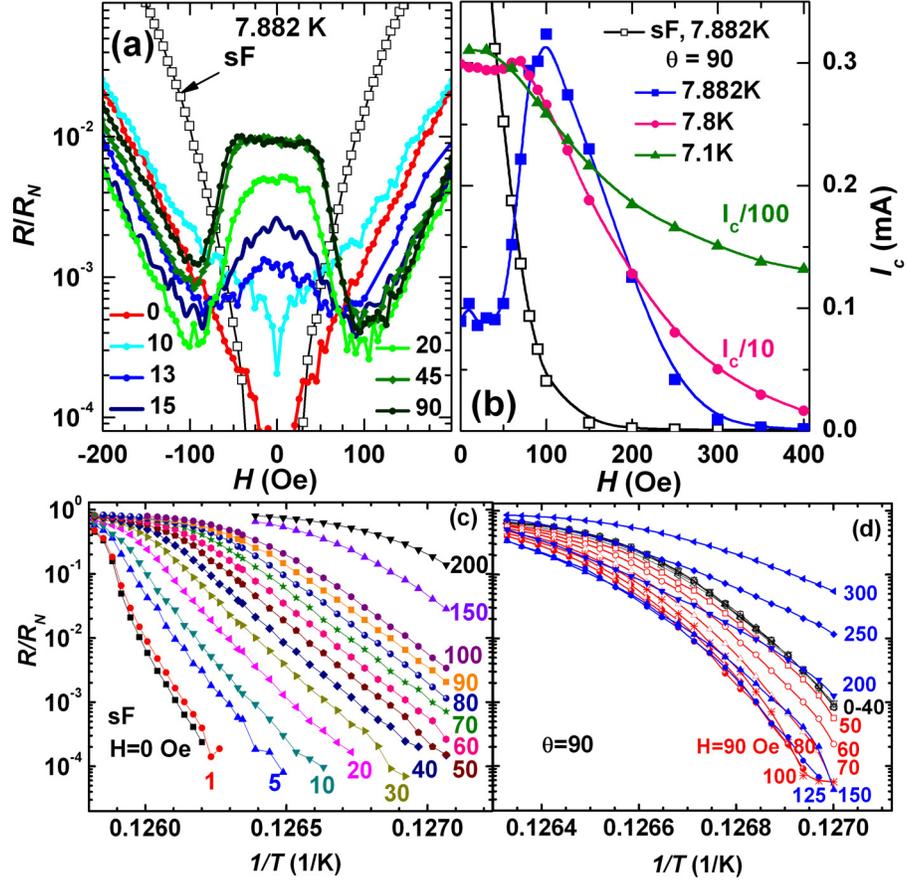

FIG. 3: (a) $R/R_N$ versus $H$ at 7.882 K for sF state, and various $\theta$. (b) Critical current $I_c$ versus $H$ for sF state at 7.882 K (open squares), and for $\theta = 90°$, $I_c$ at 7.882 K (blue), $I_c/10$ at 7.8 K (pink), and $I_c/100$ at 7.1 K (green). (c) $R/R_N$ versus $1/T$ for sF state at labeled magnetic fields $H$ = 0 to 200 Oe, and for (d) $\theta = 90°$, 0 to 40 Oe (black), 50 to 100 Oe (red), and 125 to 300 Oe (blue).

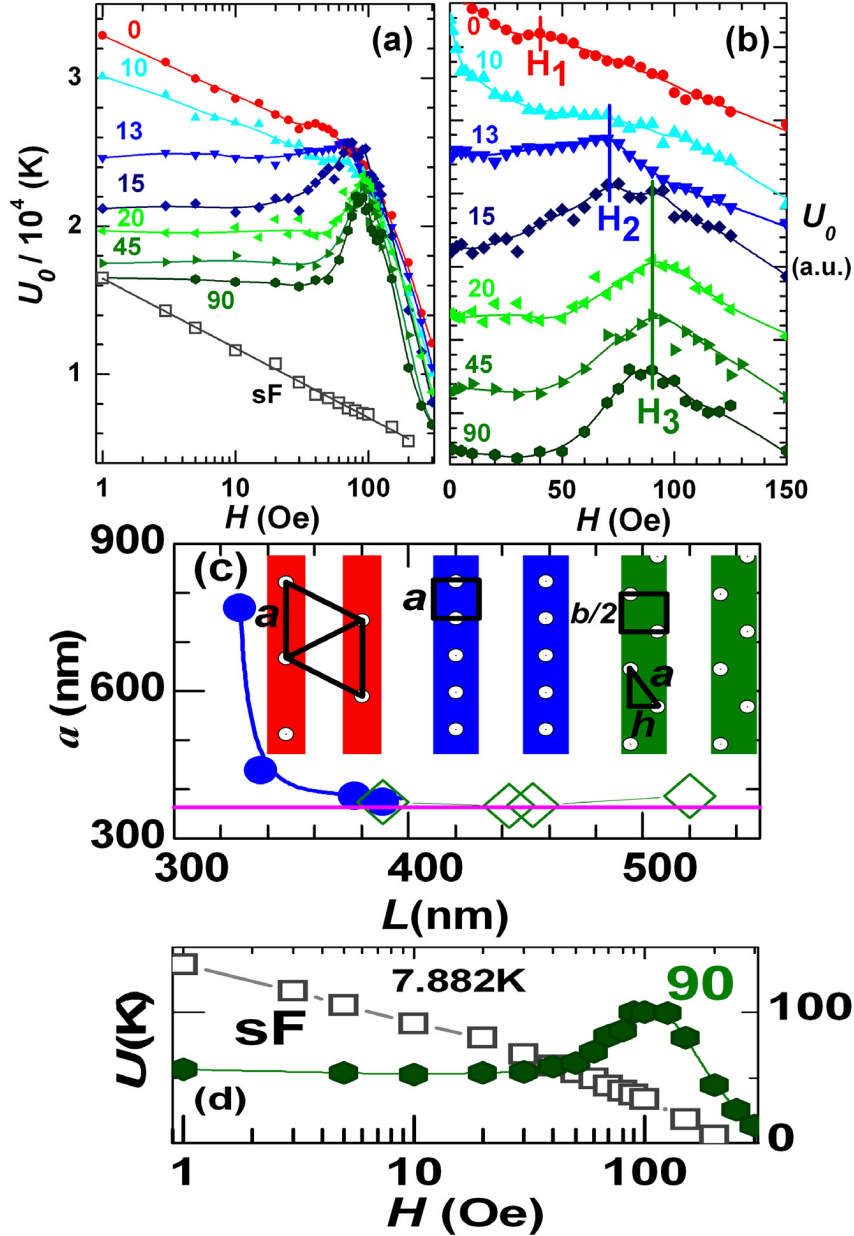

FIG. 4: (a) Activation energy $U_0$ at 0 K versus (a) $\ln H$ (b) $H$ (shifted vertically for clarity) showing peaks at $H_1$, $H_2$ and $H_3$ for sF and different $\theta$, (c) vortex separation $a$ vs. $L$ and the schematic vortex structures, for $\theta=0°$ (red), $\theta=15°$ at $H_2$ (blue), and $\theta=15°$ at $H_3$ (green); the horizontal line is that for Abrikosov lattice, (d) activation energy $U(H)$ at 7.882 K for sF and $\theta = 90°$.